\documentclass{article}

\usepackage{pdfpages}
\usepackage{multirow}
\usepackage{graphics}
\usepackage{amsmath}
\usepackage{afterpage}

\usepackage[breaklinks=true]{hyperref}

\begin{document}
\date{}

\title{Random Forest DBSCAN for USPTO Inventor Name Disambiguation\footnote{A shorter version of this paper was published in JCDL \cite{Kim2016Random}, and the full version was published in IJCAI-SBD \cite{kim2016Random2}.}}

\author{
\medskip
Kunho Kim$^{\ddag}$, Madian Khabsa$^{\mathsection}$, C. Lee Giles$^{\dag\ddag}$\\
\medskip
\scalebox{0.8}{
	\begin{tabular}{cc}
		$^{\ddag}$Computer Science and Engineering & $^{\mathsection}$Microsoft Research\\
		$^{\dag}$Information Sciences and Technology & 
		One Microsoft Way \\
		The Pennsylvania State University & 
		Redmond, WA 98005, USA\\
		University Park, PA 16802, USA &  
	\end{tabular}} \\
	\scalebox{0.8}{
		\begin{tabular}{c}
			kunho@cse.psu.edu, madian.khabsa@microsoft.com, giles@ist.psu.edu
		\end{tabular}}
}
		
\maketitle

\begin{abstract}
	
	Name disambiguation and the subsequent name conflation are essential for the correct processing of person name queries in a digital library or other database. It distinguishes each unique person from all other records in the database. We study inventor name disambiguation for a patent database using methods and features from earlier work on author name disambiguation and propose a feature set appropriate for a patent database. A random forest was selected for the pairwise linking classifier since they outperform Naive Bayes, Logistic Regression, Support Vector Machines (SVM), Conditional Inference Tree, and Decision Trees.  Blocking size, very important for scaling,  was selected based on experiments that determined feature importance and accuracy. The DBSCAN algorithm is used for clustering records, using a distance function derived from random forest classifier. For additional scalability clustering was parallelized. Tests on the USPTO patent database show that our method successfully disambiguated 12 million inventor mentions within 6.5 hours. Evaluation on datasets from USPTO PatentsView inventor name disambiguation competition shows our algorithm outperforms all algorithms in the competition.
\end{abstract}

\section{Introduction}
One of the most frequent queries for digital library search system is a person name. An example is to find all relevant records of a particular person. For a patent database, users may want to find the list of patents of a certain inventor. This query can be problematic if there is no unique identifier for each person. In that case, a method must be used to distinguish between person records in the database. This is often referred to as the personal name disambiguation problem. 

There are several factors that make this problem hard. Firstly, there can be several different formats for displaying one person's name. For example, one record has the full name "John Doe", while another contains only the iniital of first name, "J. Doe". More importantly, there are some common names that many people share. We can see this problem often with Asian names. Statistics from Wikipedia\footnote{List of common Chinese surnames, in Wikipedia. \url{https://en.wikipedia.org/wiki/List_of_common_Chinese_surname}} show that 84.8\% of the population have one of the top 100 popular surnames in China, while only 16.4\% of common names in United States. For some records the first and last name is reversed, especially for certain groups that put the last name first for their full name. Lastly, typographical errors and foreign characters can also challenge disambiguation. In addition, because of the large number of records, usually millions, manual disambiguation of all records is not feasible (which is even then not perfect) and automated methods have to be used. An automatic name disambiguation algorithm typically consists of two parts. The first is a pairwise linkage classifier that determines whether each pair of records are from the same person or not \cite{winkler2014matching}. The second is a clustering algorithm, grouping records for each unique person using the classifier. 

Here, we propose to use an author name disambiguation algorithm for the patent inventor database. Our algorithm follows the typical steps of author name disambiguation, but with a newly proposed set of features from patent metadata. Having experimented with several different classifers, we use a random forest classifier to train for pairwise linkage classification and use DBSCAN for clustering for disambiguation. We use the publicly available USPTO database for testing. Recently there was an inventor name disambiguation competition for this database. Raw data is publicly available via the competition's web page\footnote{\url{http://www.dev.patentsview.org/workshop}}. This raw data contains all published US patent grants from 1976 to 2014. Although we didn't participate in the competition, we used the same training and test datasets used in the competition for evaluation. The competition's evaluation results show our algorithm to be superior to other suggested algorithms in the competition. A detailed explanation of dataset and results can be found in results section.

\section{Related Work}
Several approaches have been proposed for pairwise linkage classification using different machine learning algorithms. Han et al. \cite{han2004two} proposed two approaches using a Hybrid Naive Bayes and support vector machine(SVM) classifier. Huang et al. \cite{Huang2006efficient} used an online active SVM(LASVM) to boost the speed of SVM classifier. Song et al. \cite{song2007efficient} used probabilistic latent semantic analysis(pLSA) and Latent Dirichlet allocation(LDA) to disambiguate names based on publication content.  Treeratpituk and Giles \cite{Treeratpituk2009disambiguating} first introduced the random forest (RF) for disambiguation and showed the random forest classifier at the time to have the best accuracy compared to other machine learning based classifiers. Godoi et al. \cite{godoi2013relevance} used an iterative approach to update ambiguous linkage with user feedback. Fan et al. \cite{fan2011graph} used graph based framework for name disambiguation, and Hermansson et al. \cite{hermansson2013entity} used graph kernels to calculate similarity based on local neighborhood structure. Instead of using machine learning algorithms, Santana et al. \cite{santana2014combining} used domain specific heuristics for classification. Recently, Ventura et al. \cite{ventura2015seeing} applied a random forest classifier with agglomerative clustering for inventor name disambiguation for USPTO database.  

For clustering algorithms for disambiguation, Mann and Yarowsky \cite{mann2003unsupervised} used a simple agglomerative clustering which still had a transitivity problem. The transitivity problem occurs when there are three records \emph{a}, \emph{b}, \emph{c} and while \emph{a} matches to \emph{b}, \emph{b} matches to \emph{c}, \emph{a} does not match with \emph{c}. Han et al. \cite{han2005name} used K-spectral clustering  which had scaling issues and the K (number of clusters) was heuristically determined. To overcome those problems, Huang et al. \cite{Huang2006efficient} proposed a density-based clustering(DBSCAN) algorithm. Another is a graphical approach using conditional random fields for clustering, using Markov Chain Monte Carlo(MCMC) methods \cite{wick2012discriminative}. Recently Khabsa et al. \cite{khabsa2015online} proposed a constraint-based clustering algorithm based on DBSCAN and extended it to handle online clustering.

\section{Disambiguation Process}

Patent records have consistent metadata to that of scholarly publications. There exists title, personal information of inventors, such as name, affiliation, etc. Ventura et al. \cite{ventura2015seeing} applied author name disambiguation algorithms to patent records, showing very promising results. Our algorithm follows the same general steps of author name disambiguation. 

First we train a pairwise classifier that determines whether each pair of inventor records is same person or not. 
Second, we apply blocking to the entire records for scaling. Finally, we cluster inventor records from each block separately using the classifier learned from the previous step. 

\subsection{Training Pairwise Classifier}
Pairwise classifier is needed to distinguish whether each pair of inventor records is the same person or not. In this section we show what features are used and how we sample the training data. We compare several machine learning classifiers to find the best one for inventor name disambiguation.

\subsubsection{Selecting Features} \label{section:feature}
\begin{table}
	\centering
	\scalebox{0.9}{\begin{tabular}{|c|c|l|} \hline 
			Category&Subcategory&Features\\ \hline
			\multirow{5}{*}{Inventor}
			& First name& Exact, Jaro-Winkler, Soundex \\ 
			& Middle name& Exact, Jaro-Winkler, Soundex\\ 
			& Last name& Exact, Jaro-Winkler, Soundex, IDF\\ 
			& Suffix& Exact\\ 
			& Order& Order comparision\\ \hline
			\multirow{3}{*}{Affiliation}
			& City& Exact, Jaro-Winkler, Soundex\\ 
			& State& Exact\\ 
			& Country& Exact\\ \hline
			Co-author& Last name& \# of name shared, IDF, Jaccard\\ \hline
			Assignee& Last name& Exact, Jaro-Winkler, Soundex\\ \hline
			\multirow{2}{*}{Group}
			& Group& Exact\\ 
			& Subgroup& Exact\\ \hline
			Title& Title& \# of term shared\\ \hline
		\end{tabular}}
		\caption{Features used for the random forest}
		\label{tab:feature}
	\end{table}
	We start with the feature set used in Ventura et al. \cite{ventura2015seeing}, and test additional features that are used in author disambiguation for scholarly databases. We only kept features that had a meaningful decrease in Gini importance if they were removed. Table \ref{tab:feature} shows all features used for the random forest classifier. A detailed explanation of each term is as follows:
	\begin{itemize}
		\item Exact: Exact string match, 3 if name matches and both full names, 2 if initial matches and not both full names, 1 if initial not matches and not both full names, 0 if name not matches and both full names.
		\item Jaro-Winkler: Jaro-Winkler distance \cite{winkler1990string} of two strings. Jaro-Winkler distance is a variant of Jaro distance. Jaro disatnace $d_{j}$ of two string $s_1$ and $s_2$ is calculated as
		\begin{displaymath}
			d_j = \begin{cases}
				0 & \text{if  } n_{m}=0 \\
				\frac{1}{3}\big(\frac{n_{m}}{|s_1|} + \frac{n_{m}}{|s_2|} + \frac{n_{m}-\frac{1}{2}n_{t}}{n_{m}}\big) & \text{otherwise}
			\end{cases}
		\end{displaymath}
		where $n_m$ is number of matching characters, and $n_t$ is number of transpositions. Each character is considered as a match only if they are within distance of a half length of a longer string $-1$. Jaro-Winkler distance $d_{jw}$ of two strings are calculated using this Jaro distance $d_j$,
		\begin{displaymath}
			d_{jw} = d_j + l_{\text{prefix}}p(1-d_j)
		\end{displaymath}
		where $l_{\text{prefix}}$ is length of common prefix between two strings (up to 4 characters). $p$ is a scaling factor, we use 0.1.
		\item Soundex: Convert each string with Soundex algorithm \cite{knuth1973art} and then do an exact string match giving credit for phonetically similar strings. The basic idea of soundex algorithm is to cluster phonetically similar consonants and convert them with the group number. \{b, f, p, v\}, \{c, g, j, k, q, s, x, z\}, \{d,t\}, \{l\}, \{m,n\}, \{r\} are the 6 groups.
		\item IDF: Inverse document frequency(calculated by \# of records total/\# of records with name) of the name, to give more weight to a unique name.
		\item Order comparison: 2 if both records are first author, 1 if both records are last author, 0 otherwise.
		\item \# of name shared: number of same name shared without considering order.
		\item \# of term shared: number of shared terms appear in both titles, excluding common stop words.
	\end{itemize}
	
	\subsubsection{Selecting Samples for Training Classifier}
	The existing labeled data from the USPTO database has two challenges in that we cannot directly use all possible pairs as a training set for a classifier. First, the majority of the labeled clusters have only a single record. In the Mixture dataset, these are 3,491 clusters out of 4,956 clusters(70.44\%) and in the Common characteristics dataset 26,648 clusters out of 30,745 clusters(86.67\%) that have only a single record. Those clusters are not useful as training data because we can only get negative pairs (two records not from the same person) from them. To train a good classifier, we need data that can give both positive and negative examples. As such we removed all those clusters, and used clusters that only have more than 1 record. 
	
	Second, there were insufficient informative negative samples from the labeled datasets. Since we need to use blocking for scaling, we want to use only pairs that consist of records from same block, since pairs from different blocks are not going to be examined in the clustering process. But there were few different clusters within each block in the labeled datasets. Since there were fewer negative samples than positive samples and to avoid overfitting, we take samples from a bigger block than the actual blocking size, using \emph{first 3 characters of last name+first name initial} while actual blocking is done with \emph{full last name+first name initial}.
	
	\subsubsection{Classifier Selection}
	
	\begin{table}
		\centering
		\scalebox{0.90}{\begin{tabular}{|c|c|c|c|} \hline 
				Method&Precision&Recall&F1\\ \hline
				Naive Bayes&0.9246&0.9527&0.9384\\ 
				Logistic Regression&0.9481&0.9877&0.9470\\ 
				SVM&0.9613&\textbf{0.9958}&0.9782\\ 
				Decision Tree&0.9781&0.9798&0.9789\\ 
				Conditional Inference Tree&0.9821&0.9879&0.9850\\ 
				Random Forest&\textbf{0.9839}&0.9946&\textbf{0.9892}\\ \hline
			\end{tabular}}
			\caption{Comparison of different classification methods}
			\label{tab:classficiation}
		\end{table}  
		
		We experimented with several supervised classifiers using the proposed feature set. We tested with the mixture of two training datasets - Mixture and Common characteristics. A detailed explanation of the datasets are in the result section. Table \ref{tab:classficiation} shows the results with 4-fold cross validation. Tree-based classifiers have a higher accuracy compared to non-tree classifiers such as SVM and logistic regression. Among them, the Random Forest (RF) classifier gives the best accuracy in terms of F1 score. RF is an ensemble classifier that aggregates the votes from decision trees for classification \cite{breiman2001random}. Previous work \cite{Treeratpituk2009disambiguating} showed RF is effective for pairwise linkage classification in scholarly database. This experiments showed that for patent database RF also has the best accuracy. We trained our RF classifier with 100 trees with 5 features tried in each split. We estimated the out-of-bag (OOB) error of the RF to measure the classification quality. OOB error is known to be an unbiased estimation of test set classification error \cite{breiman2001random}. The error rates for Common characteristics and Mixture dataset were 0.05\% and 0.07\% respectively. 
		
		\subsection{Blocking}\label{section:blocking}
		\begin{table}
			\centering
			\scalebox{0.9}{\begin{tabular}{|c|c|} \hline 
					Rank&Feature\\ \hline
					1&Last name (Jaro-Winkler)\\ 
					2&First name (Jaro-Winkler)\\ 
					3&Last name (Exact)\\ 
					4&Last name (Soundex)\\ 
					5&First name (Soundex)\\ 
					6&Affiliation (Jaro-Winkler)\\ 
					7&First name (Exact)\\ 
					8&State (Exact)\\ 
					9&Middle name (Soundex)\\ 
					10&Middle name (Exact)\\ \hline
				\end{tabular}}
				\caption{Top 10 important features of the random forest with respect to the Gini decrease}
				\label{tab:top10}
			\end{table}
			
			
			The USPTO patent database consists of 12 millions of inventor mentions. Due to the limitation of physical memory, we cannot efficiently perform the clustering for the whole database. Blocking is done in preprocessing in order to solve this problem . Records are split into several blocks, based on the blocking function. The function should be carefully selected so that records from the same person are in the same block with high probability \cite{bilenko2006adaptive}. Then we perform clustering for each block. 
			
\begin{figure}
					\centering
					\includegraphics[width=0.8\textwidth]{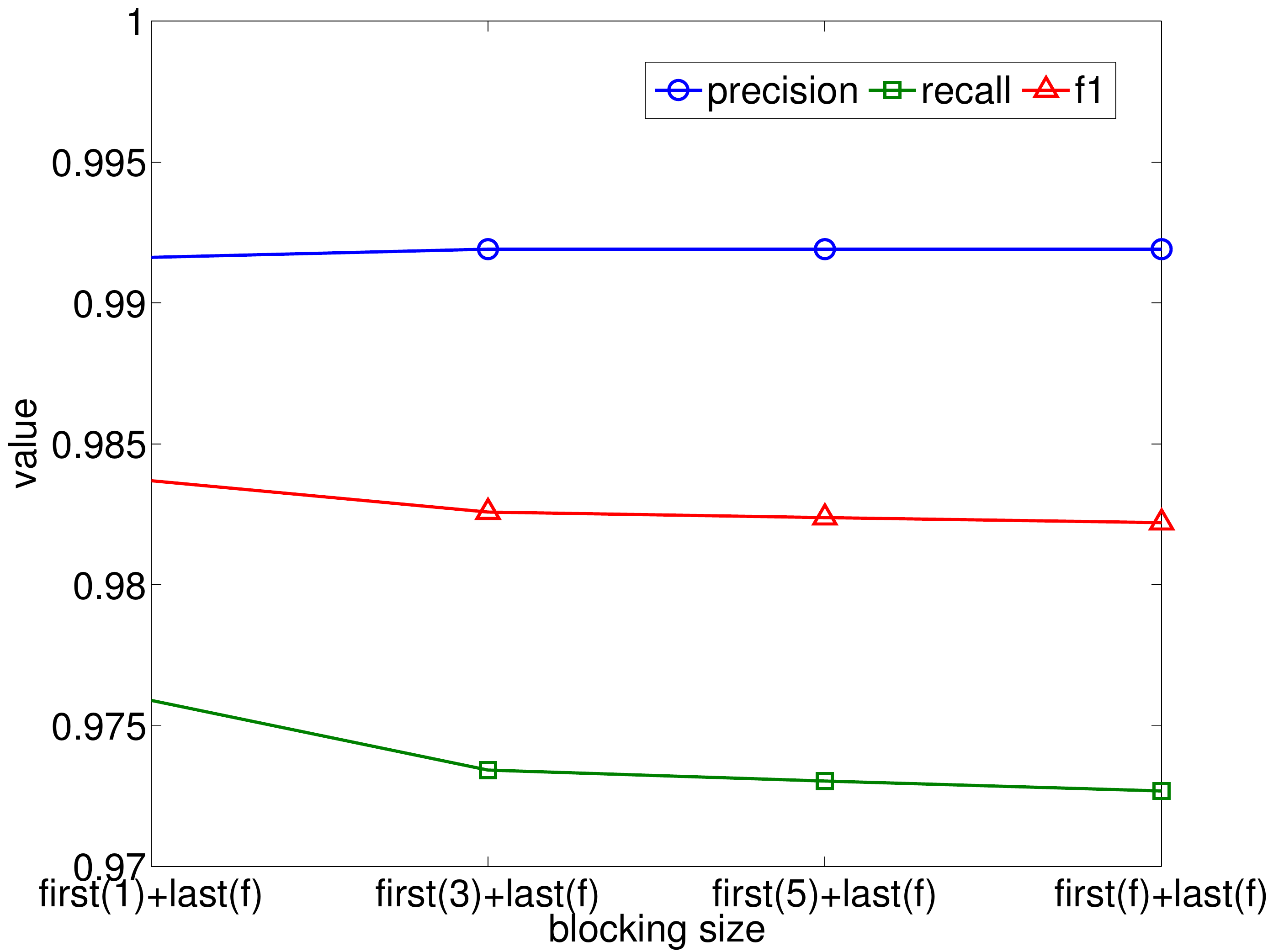} 
					\caption{Evaluation of different blocking size. FN denotes the first name and LN denotes the last name. The number (n) denotes first n characters used for blocking. If n is f, full name was used. }
					\label{fig:blocking}	
				\end{figure}

			Table \ref{tab:top10} shows the top 10 important features of the RF according to the average Gini decrease. The table shows the most important features are from first name and last name. All features from them are in top 10 except for the last name IDF. Thus, for best performance we made a blocking function with a combination of the first name and last name. Figure \ref{fig:blocking} and Table \ref{tab:blocking} shows the accuracy and computation time with respect to different block sizes. While precision is steady with different block sizes, recall gets lower as the block size becomes smaller, as does F1. This is because this blocking function splits the potential matches into different blocks. While the accuracy is getting lower, the computation time is reduced due to smaller block sizes. 
			We use \emph{full last name+initial of first name}, which was the blocking function that gives highest accuracy and that each block can be loaded fully into memory. 
			
				\begin{table}
					\centering
					\scalebox{0.9}{\begin{tabular}{|c|c|c|c|c|} \hline
							Block&FN(1)+LN(f)&FN(3)+LN(f)&FN(5)+LN(f)&FN(f)+LN(f)\\ \hline
							Time&6h 30min&5h 49min&5h 27min&5h 17min\\ \hline
						\end{tabular}}
						\caption{Computation time comparision for different block size}
						\label{tab:blocking}
					\end{table}
				
				\subsection{Clustering Using DBSCAN}
				
				We use a density-based clustering algorithm, DBSCAN \cite{ester1996density} to cluster inventor records. DBSCAN is widely used for disambiguation, because it does not require a prior the number of clusters, and it resolves the transitivity problem \cite{Huang2006efficient}.
				
				Using DBSCAN to cluster inventor records, we need to define a distance function for each pair of inventor records. The RF classifier predicts whether each pair of records are from the same person or not with a binary value(0 or 1) output. From the RF, we can get the number of negative/positive votes in its trees. We use the fraction of negative(0) votes of the trees in random forest as the distance function \cite{Treeratpituk2009disambiguating}. The final resulting clusters from DBSCAN algorithm are the result of the disambiguation. 
				
				\subsection{Parallelization} \label{section:parallelization}
				
				We use parallelization of GNU Parallel \cite{tange2011gnu} to utilize all cores available for clustering \cite{khabsa2014large}. Our work consumes memory proportional to the total number of records in the block. Due to the limitation of our physical memory, we cannot completely utilize all cores at a time if block size is too large. As such, we grouped all blocks with respect to total number of records. 
				
				The machine we use for the experiment has about 40GB memory available, and 12 cores(runs up to 24 threads simultaneously) at best. The first group consists of blocks that have less than 500 records, and we run 24 threads maximum simultaneously. The second group consists of blocks that have between 500 and 5,000 records and we run 12 threads maximum. The last group consists of blocks that have more than 5,000 records and we run 6 threads maximum. 
				
				\section{Results}\label{section:result}
				We tested our algorithm on the USPTO patent database. We used the same evaluation datasets of USPTO Patentsview inventor named disambiguation competition to compare the results. The test dataset includes ALS, ALS common, IS, E\&S, and Phase2.  
				The ALS and ALS common datasets are from Azoulay et al. \cite{azoulay2007anatomy}, which consists of inventors from the Association of Medical Colleges(AAMC) Faculty Roster. ALS common is a subset of the ALS dataset with common popular names. The IS dataset is from Trajtenberg and Shiff \cite{trajtenberg2008identification}, containing Israeli inventors in USPTO database. E\&S dataset is from Ge et al. \cite{ge2016engineer} and consists of patents from engineers and scientists. Phase2 is a random mixture of previous datasets. The training dataset includes the Mixture and Common characteristics datasets. Mixture dataset is random mixture of IS and E\&S dataset, and Common characteristics dataset is a subsample of E\&S dataset which was subsampled according to the match characteristics of the USPTO database, in terms of the mean number of inventors per patent and percentage of missing assignees. 
				
\begin{figure}
					\centering
					\includegraphics[width=0.80\textwidth]{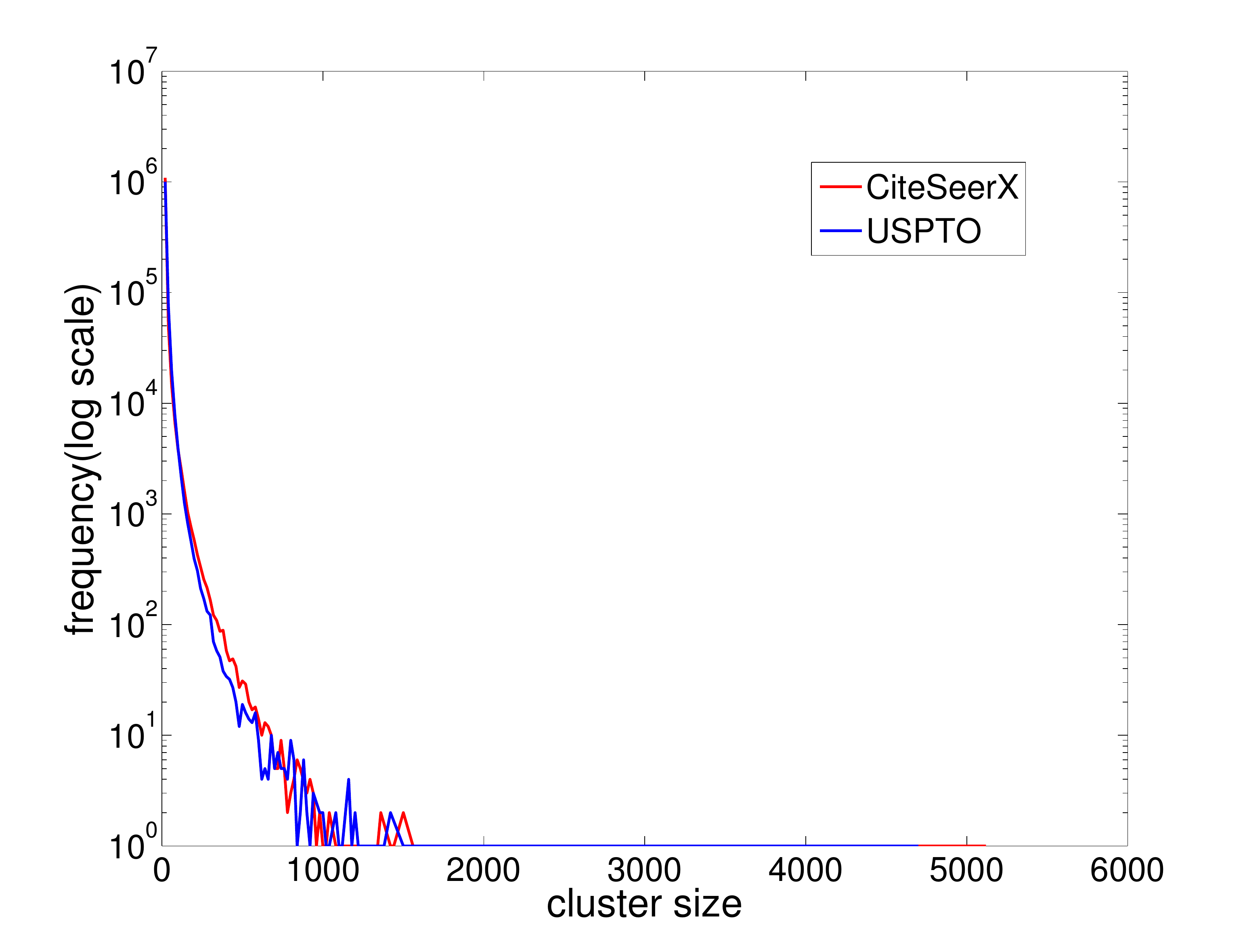} 
					\caption{Frequency of each cluster size for the CiteSeerX and USPTO database}
					\label{fig:stat}
				\end{figure}

				The result of our disambiguated USPTO database shows a similar tendency to previous disambiguation studies of scholarly databases. Figure \ref{fig:stat} shows the cluster frequency of each cluster size of CiteSeerX\footnote{\url{http://citeseerx.ist.psu.edu}} and USPTO database after disambiguation. For both databases, small clusters have high frequency and big clusters are rare with a long tail. A total of 1.11 million clusters are produced by inventor name disambiguation and the average number of patent mentions per individual inventor is 4.93. For CiteSeerX database, the average number is 6.07. 
				
				For further evaluation, we measured pairwise precision, recall, and F1 score with definitions:
				
				\begin{displaymath}\text{Pairwise Precision} = \frac{\text{\# of correctly matched pairs}}{\text{\# of all matched pairs by algorithm}}\end{displaymath}
				\begin{displaymath}\text{Pairwise Recall} = \frac{\text{\# of correctly matched pairs}}{\text{\# of pairs in manually labeled dataset}}\end{displaymath}
				\begin{displaymath}\text{Pairwise F1 Score} = 2\cdot\frac{\text{Pairwise Precision}\cdot\text{Pairwise Recall}}{\text{Pairwise Precision}+\text{Pairwise Recall}}\end{displaymath}\\
				
				
				Table \ref{tab:result} shows the results for each training and test dataset. Results were slightly better with the Common characteristics dataset, as expected from OOB error of RF. 
				This is because common characteristics dataset has more samples and is subsampled according to match the characteristics of the whole USPTO database.
				We can also see that the recall is relatively lower compare to the precision. Blocking affects the recall, as it can remove some potential matches. 
				Since we have a trade-off between efficiency and recall in our algorithm, blocking needs to be further improved for higher recall. 
				Table \ref{tab:comparison} shows F1 score comparison between our work and the best result from the competition for each test dataset. 
				The winner of the competition used a pre-defined distance metric and Markov Chain Monte Carlo(MCMC) based clustering method inspired from \cite{wick2012discriminative}. Note that our algorithm has the best performance on all datasets. The P value with one-tailed Wilcoxon test is 0.03125, which indicates that the improvement of our algorithm is statistically significant at the 0.05 level. 
				We can see from the results that the DBSCAN algorithm with the RF classifier used in scholarly disambiguation is also effective for inventor name disambiguation in a patent database.  
				
				Our disambiguation is much faster with parallelization. We used Intel Xeon X5660@2.80GHz machine with 12 cores and 40GB memory available in an idle state, configured with RHEL 6. The disambiguation process takes about 6.5 hours to finish for both training sets. Currently we cannot fully utilize all the CPUs for certain blocks that contain large number of records, because of memory limitations. 
				Better way of blocking such as \cite{bilenko2006adaptive} is needed for efficient memory usage, for fast performance and scalability.
				This remains as a future work.
				
				\begin{table}
					\centering
					\scalebox{0.9}{\begin{tabular}{|c|c|c|c|c|} \hline 
							Test Set&Training Set&Precision&Recall&F1 Score\\ \hline
							\multirow{2}{*}{ALS}
							&Mixture&0.9963&0.9790&0.9786\\ 
							&Common&0.9960&0.9848&0.9904\\ \hline
							\multirow{2}{*}{ALS common}
							&Mixture&0.9841&0.9796&0.9818\\ 
							&Common&0.9820&0.9916&0.9868\\ \hline
							\multirow{2}{*}{IS}
							&Mixture&0.9989&0.9813&0.9900\\ 
							&Common&0.9989&0.9813&0.9900\\ \hline
							\multirow{2}{*}{E\&S}
							&Mixture&0.9992&0.9805&0.9898\\ 
							&Common&0.9995&0.9810&0.9902\\ \hline
							\multirow{2}{*}{Phase2}
							&Mixture&0.9912&0.9760&0.9836\\ 
							&Common&0.9916&0.9759&0.9837\\ \hline
						\end{tabular}}
						\caption{Disambiguation evaluation}
						\label{tab:result}
					\end{table}
					
					\begin{table}
						\centering
						\scalebox{0.9}{\begin{tabular}{|c|c|c|} \hline 			
								Test Set&F1(Ours)&F1(Winner)\\ \hline
								ALS&0.9904&0.9879\\ 
								ALS common&0.9868&0.9815\\ 
								IS&0.9900&0.9783\\ 
								E\&S&0.9902&0.9835\\ 
								Phase2&0.9837&0.9826\\ \hline
								Average($\pm$stddev.)&0.9882$\pm$0.0029&0.9827$\pm$0.0035\\ \hline
							\end{tabular}}
							\caption{Comparison with the competition winner}
							\label{tab:comparison}
						\end{table}
						
						\section{Conclusions}
						
						We present a machine learning based algorithm for inventor name disambiguation for patent database. Motivated by the feature set of author name disambiguation for scholarly databases, we devised a proposed feature set that showed a significant low OOB error rate, 0.05\% at minimum. Based on experiments with several machine learning classifiers, we use random forest classifier to determine whether each pair of inventor records are from a the same inventor or not. Disambiguation is done by using DBSCAN clustering algorithm. We define distance function of each pair of inventor records as the ratio of votes in random forest classifier. In addition to make the algorithm scalable, we use blocking and parallelization, scheduling threads based on the size of blocks. Evaluation results with the dataset from USPTO PatentsView inventor name disambiguation competition shows our algorithm outperforms all algorithms submitted to the competition in comparable running time.
						
						Currently our algorithm is memory bounded since a great deal of memory is used to store inventor information needed for calculating features. This becomes a bottleneck when we parallelize the algorithm, since some of the block is huge due to the popularity of certain names. In future work, one could explore a better method for blocking for efficient memory usage. It would be interesting to see if other methods using graph or link data could be incorporated for better performance.
						
						\section{Acknowledgments}
						We gratefully acknowledge Evgeny Klochikhin and Ahmad Emad for assistance in the evaluation of the dataset used in USPTO PatentsView inventor name disambiguation competition and partial support from the National Science Foundation.

\bibliographystyle{abbrv}
\bibliography{ref}  

\end{document}